\documentclass[aps, prb, twocolumn, superscriptaddress, amsmath,  tightenlines, longbibliography]{revtex4-1}

\usepackage{dcolumn}
\usepackage{graphicx}
\usepackage{mathrsfs}
\usepackage{subfigure}
\usepackage{booktabs}
\usepackage{amsmath}
\usepackage{physics}
\usepackage{dsfont}
\usepackage{amstext}
\usepackage{amssymb}
\usepackage{amsbsy}
\usepackage{bbm}
\usepackage{amsthm}
\usepackage{graphicx}
\usepackage{color}
\usepackage[colorlinks,citecolor=blue]{hyperref}

\usepackage{url}
\usepackage[colorlinks]{hyperref}
\hypersetup{%
	plainpages=true,
	breaklinks=true,       
	hypertexnames=false,  
	pageanchor=true,
	colorlinks=true,
	linkcolor={blue},
	citecolor={red},
	urlcolor={blue},
	anchorcolor={black}
}

 \makeatletter

\newcommand{\Rmnum}[1]{\expandafter\@slowromancap\romannumeral #1@}
\makeatother

\begin{document}
\title{Effects of Disorder On Thouless Pumping In Higher-Order Topological Insulators}
\author{Congwei Lu}
\thanks{These authors contributed equally to this work}
\affiliation{Department of Physics, Applied Optics Beijing Area Major Laboratory, Beijing Normal University, Beijing 100875, China}
\author{Zhao-Fan Cai}
\thanks{These authors contributed equally to this work}
\affiliation{School of Physics and Optoelectronics, South China University of Technology,  Guangzhou 510640, China}
\author{Mei Zhang}
\affiliation{Department of Physics, Applied Optics Beijing Area Major Laboratory, Beijing Normal University, Beijing 100875, China}
\author{Haibo Wang}
\affiliation{Department of Physics, Applied Optics Beijing Area Major Laboratory, Beijing Normal University, Beijing 100875, China}
\author{Qing Ai}
\email[E-mail: ]{aiqing@bnu.edu.cn}
\affiliation{Department of Physics, Applied Optics Beijing Area Major Laboratory, Beijing Normal University, Beijing 100875, China}
\author{Tao Liu}
\email[E-mail: ]{liutao0716@scut.edu.cn}
\affiliation{School of Physics and Optoelectronics, South China University of Technology,  Guangzhou 510640, China}
\date{{\small \today}}


\begin{abstract}
	We investigate the effects of random onsite  disorder on higher-order Thouless pumping of noninteracting fermionic Benalcazar-Bernevig-Hughe (BBH) model. The interplay of disorder-induced topological phase transition and delocalization-localization transition is extensively explored. The higher-order Thouless pumping is characterized by the quantized corner-to-corner charge transport and nonzero Chern number, and the delocalization-localization transition is analyzed by utilizing both inverse participation ratio and finite-size scaling. The results show that the quantized corner-to-corner charge transport is broken for the strong  disorder, where the instantaneous bulk energy gap is closed. Although the instantaneous eigenstates are localized for the weak disorder, the charge transport remains quantized. This is attributed to delocalized Floquet states caused by the periodic driving.  Furthermore, the phase transition from the quantized charge transport to topologically trivial pumping is accompanied by the disorder-induced delocalization-localization transition of Floquet states. 
\end{abstract}

\maketitle

\section{\label{sec:introduction}Introduction}

Thouless charge pumping, proposed by Thouless in 1983 \cite{PhysRevB.27.6083}, is originally a dynamical topological effect in one-dimensional systems. It shares the same topological origin as the static Chern insulator, where one of the momentum coordinates is replaced by an adiabatically varying parameter. The Thouless charge pumping serves as one of the simplest manifestations to understand the topology in quantum systems, where the quantized charge transport reveals the correspondence between polarization in the bulk and charge at the boundary \cite{PhysRevB.27.6083, QNiu1984}, the so-called bulk-boundary correspondence. Recently, an unconventional bulk-boundary correspondence occurs in higher-order topological insulators (HOTIs) \cite{PhysRevLett.110.046404,Benalcazar61,PhysRevB.96.245115,PhysRevLett.119.246401,PhysRevLett.119.246402,PhysRevB.97.241405,peterson2018quantized,serra2018observation,arXiv:1801.10053,PhysRevLett.120.026801,schindler2018higher,Zhang2019,Ni2018,xue2019acoustic,arXiv:1801.10050,arXiv:1802.02585,PhysRevLett.123.216803,PhysRevResearch.2.033029,PhysRevLett.124.036803,PhysRevB.101.241104,PhysRevLett.124.063901,PhysRevB.103.L201115,arXiv:2202.12151,PhysRevLett.126.066401,PhysRevB.92.085126,PhysRevB.104.134508}: a $d$-dimensional $n$th-order ($n \geq 2$) topological system hosts topologically protected  gapless states on its $(d - n)$-dimensional boundaries. Correspondingly, the higher-order Thouless charge pumps, which can be utilized to reveal such an unconventional bulk-boundary correspondence, have been put forward and studied \cite{Wienand2022,PhysRevResearch.2.022049,PhysRevB.105.195129,PhysRevA.106.L021502,Wu_2022, arXiv:2208.08625}. Especially, the higher-order topological pumped corner-to-corner charge flow has been related to four higher-order Zak phases in a square  Boson-Hubbard model \cite{Wienand2022}, providing us a simple theoretical framework for  characterizing higher-order Thouless pumping.

Disorder, ubiquitous in real materials, plays an important role in quantum transport due to the nontrivial interplay of disorder-induced Anderson localization \cite{Lagendijk2009} and topological Anderson insulator \cite{PhysRevLett.102.136806}.  Recently, many efforts have been devoted to understanding the robustness properties of quantized charge transport against disorder in the conventional Thouless pumping \cite{PhysRevLett.121.126803,physRevLett.123.26601,PhysRevA.103.043310,Nakajima2021,Cerjan2020}, where the breakdown of quantized charge transport has been linked to a delocalization-localization transition coincident with the topological transition \cite{physRevLett.123.26601}. Recently, the disorder-induced higher-order topological pumping (i.e., topological Anderson-Thouless pump) is explored \cite{arXiv:2208.08625}. Meanwhile, the effects of the disorder on Thouless pumping in higher-order topological insulators still remains largely unexplored. 

In this paper, we aim to study the topological phase transition and delocalization-localization properties of Thouless pumping against disorder in a higher-order topological insulator. We introduce  random onsite disorder into the noninteracting fermionic Benalcazar-Bernevig-Hughe (BBH)  model \cite{Benalcazar61} under periodic driving. The quantized corner-to-corner charge transports through each corner during one topological driving period is linked to four higher-order Zak phases and characterized by the winding of the higher-order Zak phase (i.e., Chern number). Results have shown that the disorder causes the topological phase transition from Thouless pumping to trivial pumping.  Remarkably, the higher-order Thouless pumped charge transport survives despite localized instantaneous eigenstates for the weak disorder. The inverse participation ratio indicates that the quantized corner-to-corner charge transport is related to the delocalized Floquet states in the weak disorder strength. Furthermore, as the disorder strength varies, the phase transition from topological pumping to trivial pumping and delocalization to localization transition coexist in our investigated system.

The rest of the paper is structured as follows. In Sec.~\ref{sec:the model}, we describe the disordered BBH model under periodic driving. In Sec.~\ref{Thouless Pumping }, we investigate the corner-to-corner charge transport under the higher-order Thouless pumping and relate it to the four higher-order Zak phases. The topological phase transition  and delocalization-localization properties of higher-order Thouless pumping against disorder are studied in Sec.~\ref{disorder }. Finally, we conclude the work in Sec.~\ref{sec:Conclusion}.

\begin{figure}[!tb]
	\centering
	\includegraphics[width=8.4cm]{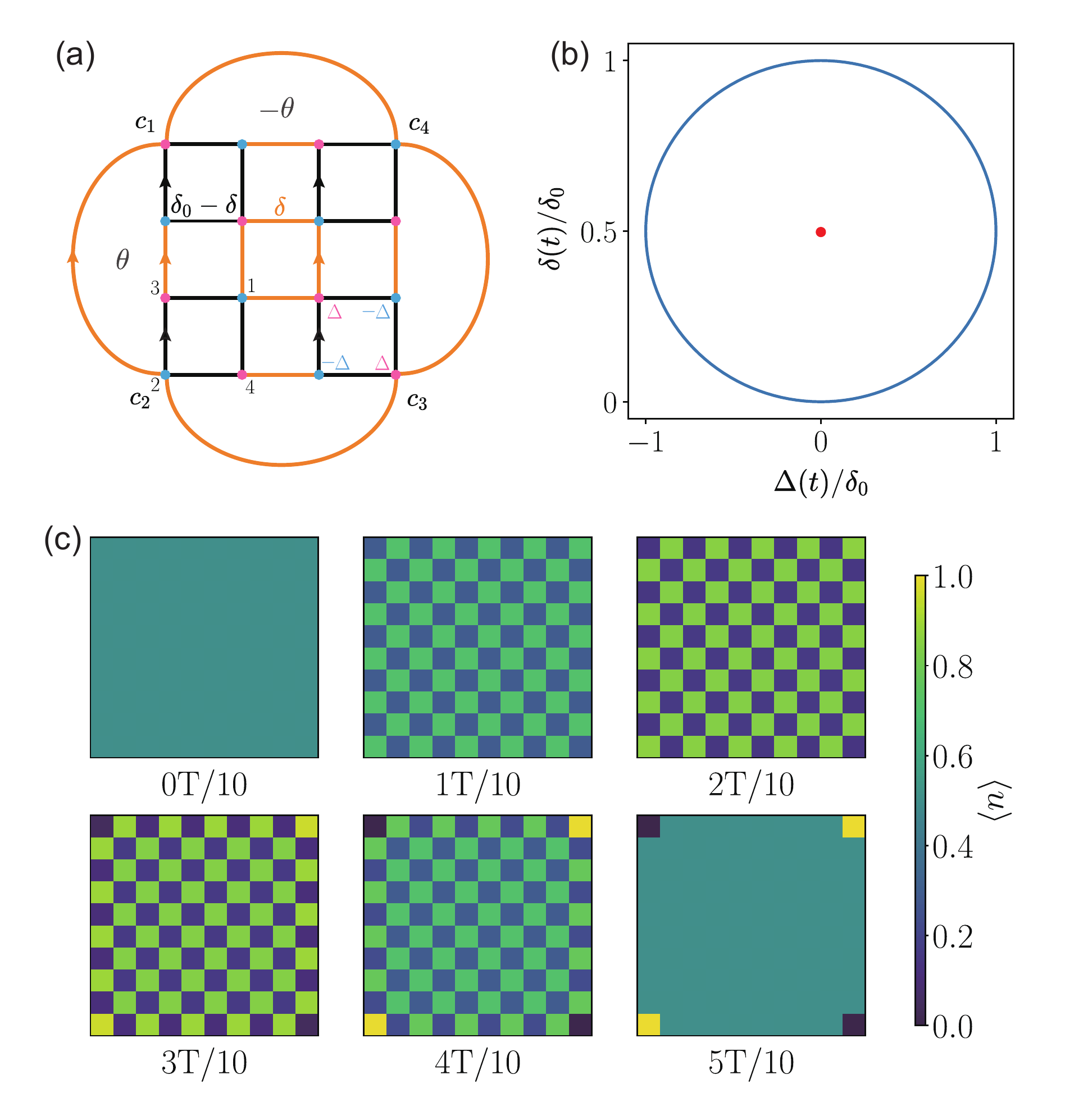}
	\caption{Higher-order topological  pumping in a noninteracting fermionic BBH model. (a) Schematic representation of a square lattice structure in the presence of time-dependent staggered nearest-neighbor hopping strength with $1-\delta$ and $\delta$,  and onsite potentials with $\Delta$ and $-\Delta$. The arrows indicate the nearest-neighbor hopping with a negative sign, accounting for a flux of $\pi$ threading each plaquette. Under corner periodic boundary conditions, $\theta$ denotes the twisted phase added to a pair of corner-connecting links with a shared corner $c_{1}=(-D, D)$. This is equivalent to adiabatically inserting a magnetic flux into the two supercells meeting at corner $c_1$.  (b) The trajectory of driving parameters $\delta(t)$ and $\Delta(t)$ in a pumping cycle. The quantized topological pumping requires such a trajectory enclosing a  gapless point (red dot) of the system's eigenenergies in the driving parameter space. (c)  Evolution of  density during a half-Thouless pump with $t\in[0, ~T/2]$. The system is initially prepared in the half-filling ground state of  $\mathcal{H}_0(t=0)$ with $\delta = 0$, and $\Delta = 0$.  At the end of the half-cyclic evolution, there are $+1/2$ charges pumped to two anti-diagonal corners and $-1/2$ charges at two diagonal corners. The parameter used in the simulation is $\hbar \omega=0.05\delta_0$. }\label{Fig1}
\end{figure}

\section{\label{sec:the model}The Model }

We extend the spinless fermionic BBH model \cite{Benalcazar61}   to include the onsite staggered potential and periodic driving on a square lattice with  $N/2 \times N/2$ unit  cells, as shown in Fig.~\ref{Fig1}(a). The system Hamiltonian reads
\begin{align}\label{MIH1}
\mathcal{H}_0 = ~&-\left[\sum_{x=-D}^{D-1} \sum_{y=-D}^{D} \left(\lambda(x)\hat{a}_{x+1,y}^\dagger \hat{a}_{x,y}+\textrm{H.c.}\right)+x \leftrightarrow y\right]  \nonumber \\
& + \Delta \sum_{x,y=-D}^{D} \hat{n}_{x,y} (-1)^{x+y} ,
\end{align}
with $\lambda(\varepsilon)$, and $\varepsilon \in \{x,~y\} $   satisfying
\begin{align}\label{MIH12}
	\lambda(\varepsilon) = \begin{cases}
		\delta_0 - \delta, ~~~~ \varepsilon \in \{-D, -D +2 \cdots, D-1\},  \\
		\delta, ~~~~ \varepsilon \in \{-D+1, -D +3 \cdots, D-2\},  \\
	\end{cases}
\end{align}
where $D=(N-1)/2$,  $\hat{a}_{x,y}^\dagger$ creates a fermion at site $(x,~y)$, $\hat{n}_{x,y}$ denotes the fermion number, $\delta$ and $\delta_0-\delta$ , with $\delta=\delta_0(1-\cos\omega t)/2$, represent the time-dependent staggered nearest-neighbor hopping strengths, and $\Delta $ and $-\Delta $, with $\Delta = \delta_0\sin \omega t$, are the time-dependent staggered onsite potentials. The undriven BBH model exhibits a second-order topological phase, featured by the appearance of in-gap states and fractional charge on corners of the lattice \cite{Benalcazar61}. The periodically driven fermionic BBH model can be experimentally realized using  ultracold quantum gases \cite{PhysRevLett.107.255301,PhysRevLett.111.185301,Goldman2016,Nakajima2016,Lohse2015}.

\section{ Higher-order Thouless Pumping and Topological Invariant}\label{Thouless Pumping }

A Thouless pump induces a quantized amount of charge transport during an adiabatic cycle in the parameter space. We now consider a half-cyclic adiabatic pump during $t\in[0,~T/2]$, with $T$ being the pumping period, as shown in Fig.~\ref{Fig1}(b). We initialize  the system in the half-filling ground state of $\mathcal{H}_0(t=0)$ with $\delta = 0$, and $\Delta = 0$. Then, the   particle density in site $i$ at time $t$ is given by 
\begin{align}\label{APNE}
	\langle \hat{n}_{i}(t) \rangle=\sum_{\alpha=1}^{N^{2}/2}\left|\bra{x_{i}}\hat{\mathcal{U}}(t)\ket{\psi_{\alpha}}\right|^{2},
\end{align} 
where $\hat{\mathcal{U}}(t)$ is the one-body evolution operator, $\ket{\psi_{\alpha}}$ is the $\alpha$-th  occupied eigenvector  of initial Hamiltonian $\mathcal{H}_0$ at $t=0$ in the bases of one-particle Fock state   and $\ket{x_{i}}$ denotes the coordinate eigenvector with eigenvalue $x_i$, which is the coordinate of site $i$. 

The half-cyclic adiabatic evolution of the particle density is shown in Fig.~\ref{Fig1}(c).  The direction of charge transport is controlled by the onsite potential $\pm\Delta$. Initially, the density is  homogeneously distributed at each site for $\delta = 0$, and $\Delta = 0$. As time evolves, when $\phi=\omega t < \pi/2$,  $\Delta$  increases, and the system is in the topologically trivial phase. The density accumulates towards sites with negative onsite potential $-\Delta$. Once $\phi$ exceeds $\pi/2$, $\Delta$ decreases, and the system enters the topological phase regime. The density, accumulated at sites with negative onsite potential, begins reducing in the bulk and along the edge, and finally becomes uniform distribution at the end of the half-cyclic evolution. However, there are $+1/2$ charges pumped to two anti-diagonal corners and $-1/2$ charges at two diagonal corners at the end of the half-cyclic evolution.

The higher-order topological charge pumping can be characterized by the four Chern numbers \cite{Wienand2022}, each of which is defined as the winding number of four higher-order Zak phases in the square-lattice geometry. The higher-order Zak phase is introduced under the corner periodic boundary conditions (CPBCs) \cite{Wienand2022} by applying corner-connecting links, with its Hamiltonian reading
\begin{align}\label{CPBC1}
\mathcal{H}^{\text{C}}_{i}(\theta) = e^{-i \theta \hat{n}_{c_i}} \mathcal{H}^{\text{C}}e^{i \theta \hat{n}_{c_i}},
\end{align}
where the corner-connecting link Hamiltonian is written as
\begin{align}\label{CPBC2}
	\mathcal{H}^{\text{C}}  =-\delta\left( \hat{a}_{c_{1}}^{\dagger} \hat{a}_{c_{2}}+ \hat{a}_{c_{2}}^{\dagger} \hat{a}_{c_{3}}   +  \hat{a}_{c_{3}}^{\dagger} \hat{a}_{c_{4}}- \hat{a}_{c_{4}}^{\dagger} \hat{a}_{c_{1}}+\text{H.c.}\right),
\end{align}
with  corner sites  $c_{1}=(-D,D)$, $c_{2}=(-D,-D)$, $c_{3}=(D,-D)$ and $c_{4}=(D,D)$. In Eq.~(\ref{CPBC1}),  $\theta$ denotes the twisted phase added to a pair of corner-connecting links with a shared corner $c_i$ [see Fig.~\ref{Fig1}(a)]. 
   
With the total Hamiltonian  under the corner periodic boundary conditions $\mathcal{H}^{\textrm{CPBC}}_{i}=\mathcal{H}_{0}+\mathcal{H}^{\text{C}}_{i}$, we can define the  high-order Zak  phase $\gamma_{i}$ ($i=1,2,3,4$) \cite{Wienand2022} as   
\begin{align}\label{HighorderZakPhase}
\gamma_{i}=i\int_0^{2 \pi} d \theta\left\langle\Psi_i(\theta)\left|\partial_\theta\right| \Psi_i(\theta)\right\rangle,
\end{align}
where the wave function $\ket{\Psi_{i}(\theta)}$ is the instantaneous   many-body  ground state of $\mathcal{H}^{\textrm{CPBC}}_{i}$ at half filling.  
By defining the current operator, $\hat{\mathcal{J}}_i = -\partial_{\theta} \mathcal{H}^{\textrm{CPBC}}_i|_{\theta=0}$, associated to the flux $\theta$ inserted into the corner-connecting links, one can infer that each of four higher-order Zak  phases is related to the charge transport $Q_{c_i}$ across the corner  $c_{i}$ along a diagonal direction  \cite{Wienand2022}
\begin{align}\label{Eq6}
	Q_{c_i}=-\frac{\Delta \gamma_i}{2 \pi},
\end{align}
where $\Delta \gamma_i$ is the gauge-invariant difference of $\gamma_{i}$ during an adiabatically pumping cycle.

Numerically, the total amount of transported charge    during a Thouless pumping circle can be calculated as
\begin{align}\label{Eq7}
	Q_{c_i}= \int_0^T \mathrm{~d} \tau \operatorname{tr}\left(\hat{\rho}(\tau) \hat{\mathcal{J}}_i\right),
\end{align}
where $\hat{\rho}(\tau)$ is the density matrix of the adiabatic evolution  \cite{Wienand2022} . Furthermore, according to Eqs.~(\ref{HighorderZakPhase}) and (\ref{Eq6}), the total charge pumped to each corner along the diagonal direction over one pumping period $T$ can be characterized by one of four Chern numbers
\begin{align}\label{Eq8}
\mathcal{C}_{i}=&\frac{1}{2\pi}\int_{0}^{T}\mathrm{~d} \tau \partial_{\tau}\gamma_{i} = -Q_{c_i},
\end{align} 
which is always quantized. Therefore, the corner-to-corner charge transport  in higher-order topological pumping is integer quantized. Note that $\sum_{i=1}^4 \mathcal{C}_{i} =0$ due to net charge conservation.

\begin{figure}[!tb]
	\centering
	\includegraphics[width=8.4cm]{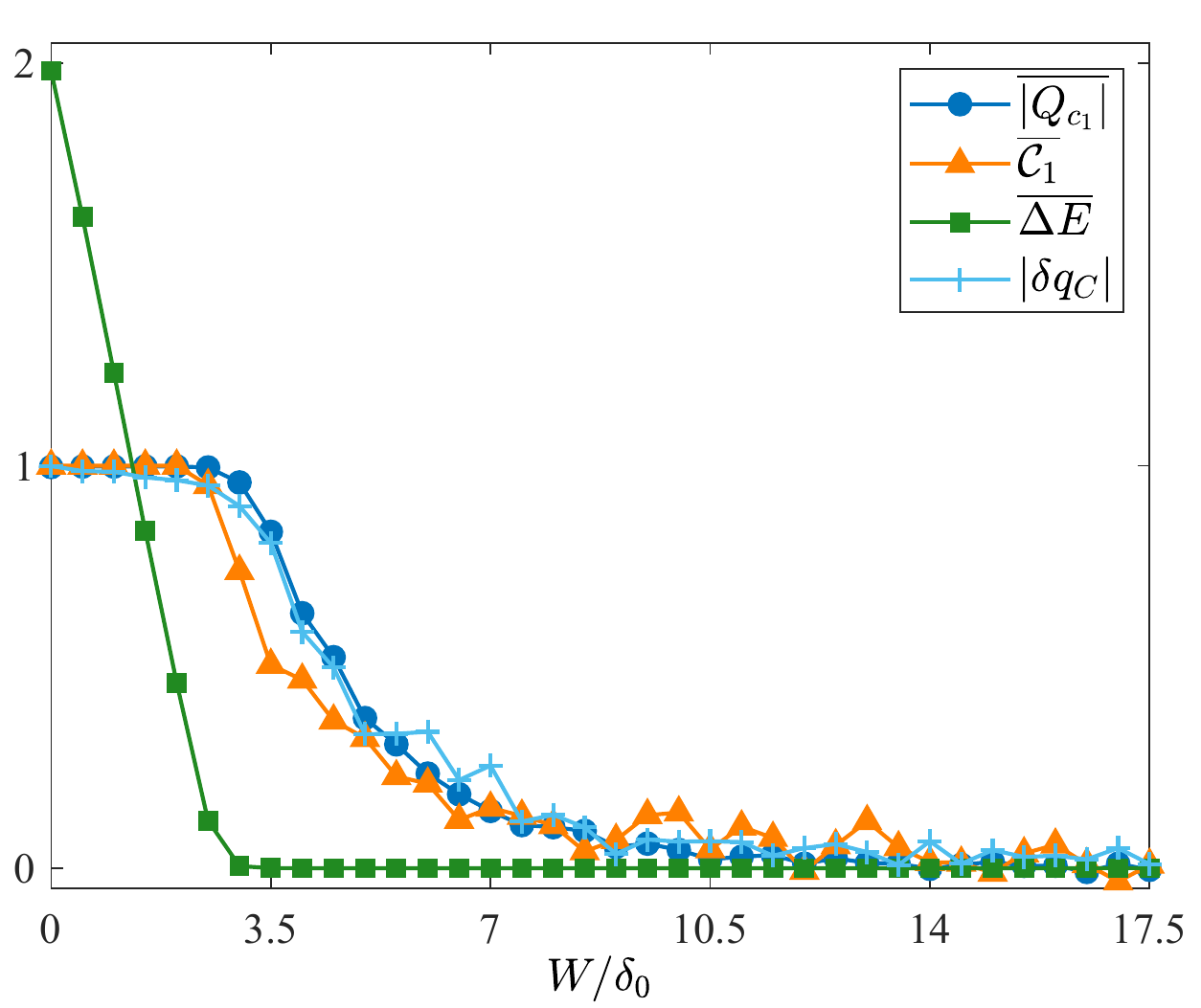}
	\caption{The disorder-averaged  corner charge $\overline{|Q_{c_1}|}$ transported across the corner $c_1$, Chern number $\overline{\mathcal{C}_1}$, and minimum value $\overline{\Delta E}$ of the instantaneous eigenenergy gaps  during a pumping cycle as a function of the disorder strength $W$ under the corner periodic boundary conditions. $\delta q_c$ is the disorder-averaged change of corner charge across the corner $c_1$, calculated under the open boundary conditions, in a pumping cycle. The results are averaged over 800 disorder realizations, with the lattice size $N\times N=24\times 24$, $\hbar \omega=0.02 \delta_0$. The other results are averaged over 600 disorder realizations, with the lattice size $N\times N=16\times16$, $\hbar \omega=0.05 \delta_0$. The breakdown of the quantized corner charge transport  coincides with the minimum instantaneous energy gap closing induced by the strong disorder strength along the pumping cycle, where the adiabatic approximation is broken.}\label{Fig2}
\end{figure}

\begin{figure*}[!tb]	
	\centering
	\includegraphics[width=1\linewidth]{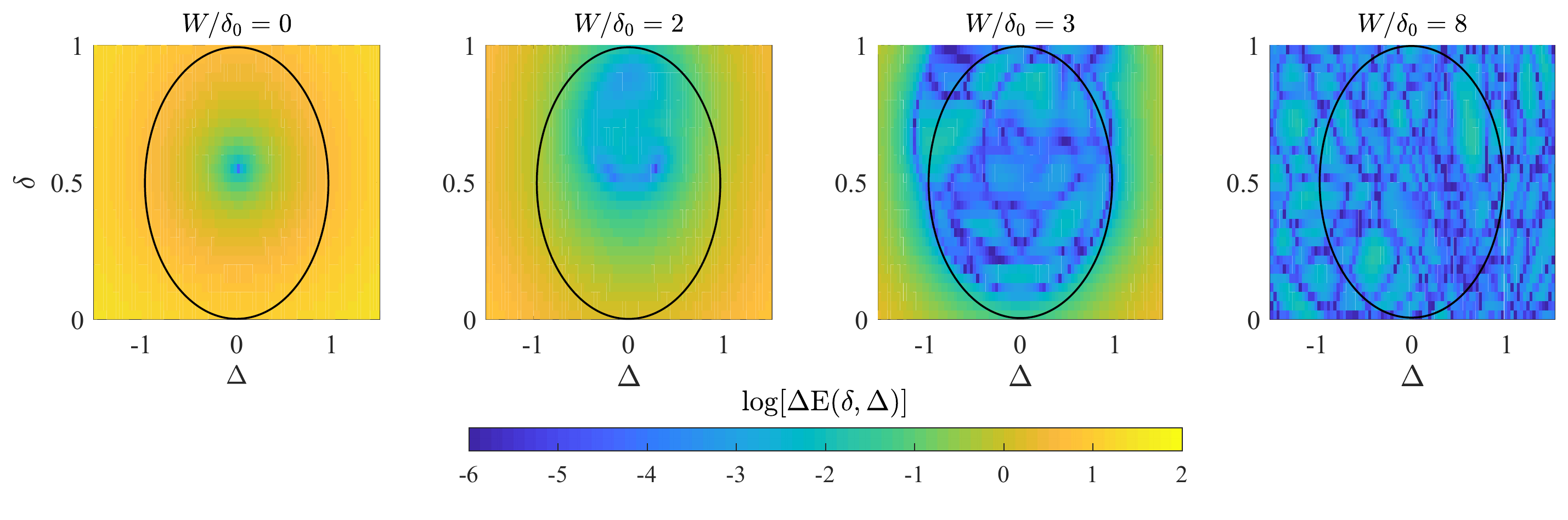}
	\caption{Gap structure $\log[\Delta E(\delta,\Delta)]$ at different disorder strengths for a single random realization. The black circle indicates the pumping cycle in the $\Delta-\delta$ parameter space.} \label{Fig3}
\end{figure*}

\section{  EFFECTS of disorder ON Higher-Order THOULESS PUMPING}\label{disorder }

\subsection{Charge transport versus disorder strength}

The main focus of this work is to explore the effects of the  disorder on higher-order Thouless pumping. We consider introducing the onsite disorder into the pumping system, which has the Hamiltonian $	  \mathcal{H}_{\textrm{tot}} = \mathcal{H}_0 + \mathcal{H}_\textrm{dis} $, with the disorder term being
\begin{align}\label{MIH3}
\mathcal{H}_\textrm{dis} = \sum_{x,y=-D}^{D} V(x,y),
\end{align}
where $V(x,y) = W \zeta_{x,y} \hat{n}_{x,y}$ denotes the onsite disordered potential, with $W$ being the disorder strength, and $\zeta_{x,y} \in [-1/2,~ 1/2]$ being uniformly
distributed random number.

Figure \ref{Fig2} shows the disorder-averaged corner charge  $\overline{|Q_{c_1}|}$, transported across the corner $c_1$,  as a function of the disorder strength $W$ in a pumping cycle under the corner periodic boundary conditions. Quantized corner-to-corner charge transport, with $\overline{|Q_{c_1}|} = 1$, persists  for the weak disorder strength with $W \lesssim 3\delta_0$. As the disorder strength further increases, $\overline{|Q_{c_1}|}$ decreases and becomes zero for the strong disorder with  $W \gtrsim 10\delta_0$. In addition, we calculate  the disorder-averaged change $\delta q_c$ of corner charge across the corner $c_1$, calculated under the open boundary conditions, in a pumping cycle (see Fig.~\ref{Fig2}). Its variation follows $\overline{|Q_{c_1}|}$ as the disorder strength is increasing. Note that $\delta q_c$ is counted by simply summing the particle density near the corner $c_1$ during a pumping cycle, where the system is initialized in   topological corner states. 

The breakdown of quantized charge transport at the strong disorder is also revealed by the disorder-averaged Chern number with $\overline{\mathcal{C}_1} = 0$ (see Fig.~\ref{Fig2}).  As the disorder strength varies, the  behavior of   Chern numbers  remains the same as that of the passing corner charge despite the existence of disorder, indicating that the Chern number, defined as the winding of higher-order Zak phase, can characterize the higher-order topological pumping in the disordered system. 

Due to its topological nature, the quantized Thouless pumping requires the trajectory of   driving parameters enclosing a gapless point of the system's eigenenergies along the pumping cycle while a gapped energy gap remains [see Fig.~\ref{Fig1}(b)]. However, the eigenenergy gap can close during the adiabatic evolution along the trajectory in the driving parameter space under  effects of the disorder, leading to the breakdown of the quantized corner-to-corner charge transport. We plot the disorder-averaged minimum value $\overline{\Delta E}$ of the instantaneous eigenenergy gaps in a pumping cycle as the disorder strength varies in Fig.~\ref{Fig3}, where $\Delta E$ is defined as 
\begin{align}\label{Eq11}
\Delta E = \min _{\phi \in[0, 2\pi]}\left[E_{N^{2}/2+1}(\phi)-E_{N^{2}/2}(\phi)\right],
\end{align}
with $E_N(\phi)$ being the $N$-particle many-body ground-state eigenenergy. As shown in Fig.~\ref{Fig2},  the breakdown of the quantized corner charge transport  coincides with the minimum instantaneous energy gap closing induced by the strong disorder strength along the pumping cycle, where the adiabatic approximation is broken.

The effects of the disorder on the quantized corner-to-corner charge transport can also be revealed by calculating $\log[\Delta E(\delta,\Delta)]$ at different disorder strengths, where $\Delta E(\delta,\Delta)$ is the instantaneous energy gaps in the $\delta-\Delta$ parameter space. As shown in Fig.~\ref{Fig3}, the black circle indicates the pumping cycle in the $\Delta-\delta$ parameter space. For $W=0$, the driving circle encircles a single gapless point (indicated by the blue point in Fig.~\ref{Fig3}), signaling a quantized higher-order topological charge transport. When the disorder strength rises to $W=2\delta_0$, the number of gapless points, with the $\Delta E$ closing to zero for these points (also $\Delta E \ll \omega$), increases. These points form a net-line region. However, the net-line region can still be encircled by the driving circle, and the quantized transport can be sustained. When the disorder strength is further increased, the net lines (indicated the blue regime) start to cross with the driving circle. While, for $W=3\delta_0$, we can enlarge the driving cycle to restore the quantized transport. However, for the strong disorder (e.g., $W=8\delta_0$), the net lines, connected by gapless points, cover  the whole parameter space, and the quantized charge transport is completely broken.

\subsection{State localization and delocalization }

We now proceed to investigate state localization and delocalization  under   the effects of the disorder in a pumping cycle, which determines the quantized charge transport. According to Anderson localization \cite{Lagendijk2009}, the arbitrary disorder can cause one- and two-dimensional states to be localized  in a non-interacting stationary system. We, therefore, explore the   localization-delocalization properties of eigenstates in the pumping circle under the corner periodic boundary conditions.  

To measure the degree of state localization, we utilize the inverse participation ratio (IPR) $I$ \cite{JTEdwards_1972}, which defines as
\begin{align}\label{Eq12}
	I(\ket{\psi_n(\phi)}) = \sum_{j=1}^{N^2}|\bra{j} \ket{\psi_n(\phi)}  |^4,
\end{align}
where $\ket{\psi_n(\phi)}$ is the $n$-th single-particle instantaneous eigenstate at pump angle $\phi$, and $\ket{j}$ is the coordinate eigenstate localized at site $j$. The IPR $I$ can vary from $I=1$ for the perfect localization to $I=1/N^2$ for the totally delocalized situation  (i.e., take $|\bra{j} \ket{\psi_n(\phi)}  | = 1/N$). Then, we can calculate the characteristic localization length by using the inverse of the minimum IPR over all eigenstates and pump angles $\phi$, which is defined as  
\begin{align}\label{Eq13}
	\xi_{\textrm{I}}=\{\min _{n,\phi}\bar{I}(\ket{\psi_n(\phi)})\}^{-1/2}.
\end{align}

\begin{figure}[!tb]
	\centering
	\includegraphics[width=8.4cm]{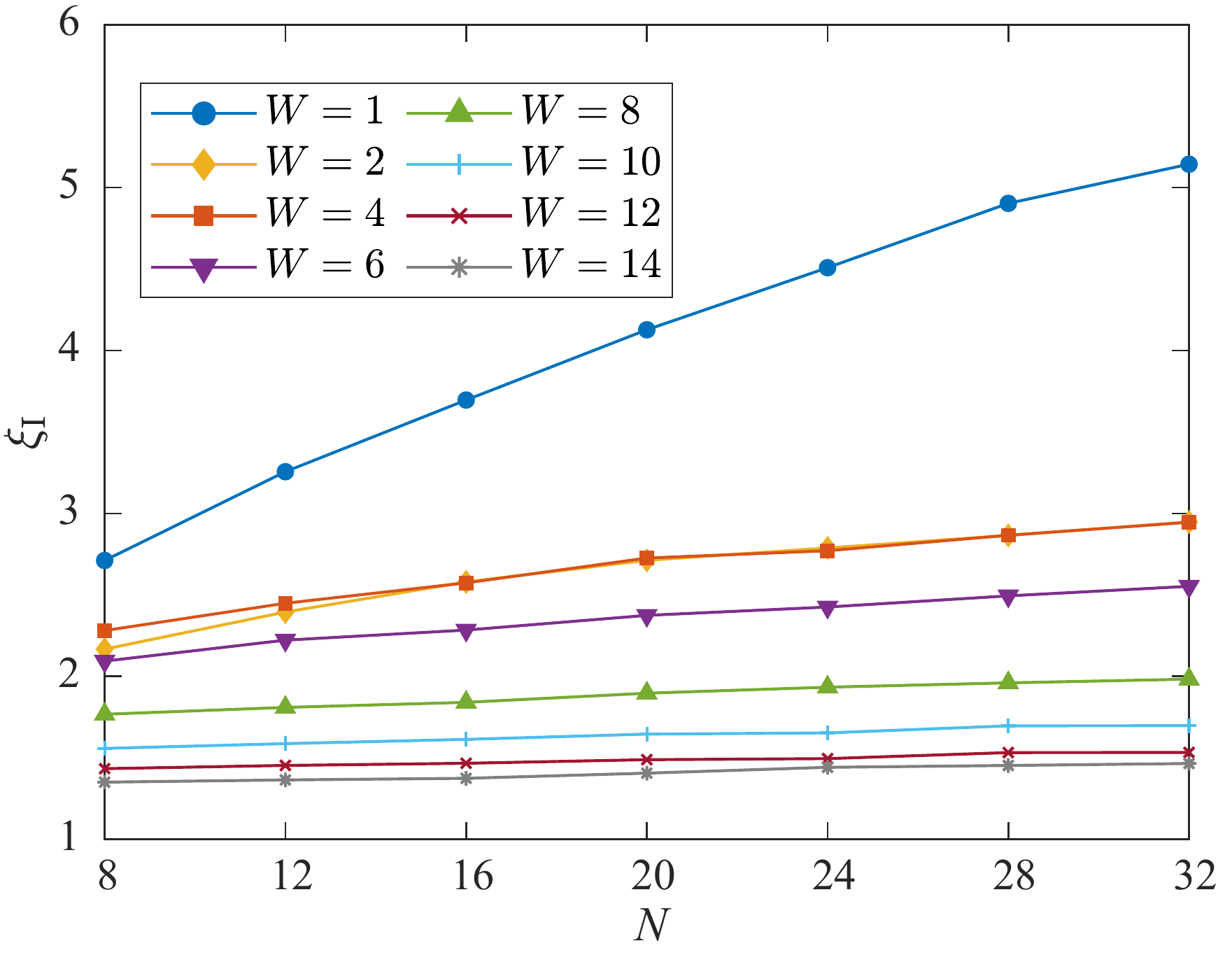}
	\caption{The characteristic localization length $\xi_{\textrm{I}}$ of the instantaneous eigenstates versus the system size $N$ for different disorder strengths. The results are averaged over 1200 disorder realizations for $N\leq 16$, over 800 disorder realizations for $24\geq N>16$, and over 400 disorder realizations for $N>24$. When $W \geq 2$, the finite size effect can be neglected. The results indicate that instantaneous eigenstates are localized even for the weak disorder. }\label{Fig4}
\end{figure}

In Fig.~\ref{Fig4}, we plot the characteristic localization length $\xi_{\textrm{I}}$ of instantaneous IPR versus the system size $N$ for different disorder strengths. When $W \geq 2$, $\xi_{\textrm{I}}$ remains nearly unchanged, and  the finite size effect can thus be neglected. Furthermore, the results show that instantaneous eigenstates are localized even in the small disorder strength.
  
\begin{figure}[!tb]
	\centering
	\includegraphics[width=8.4cm]{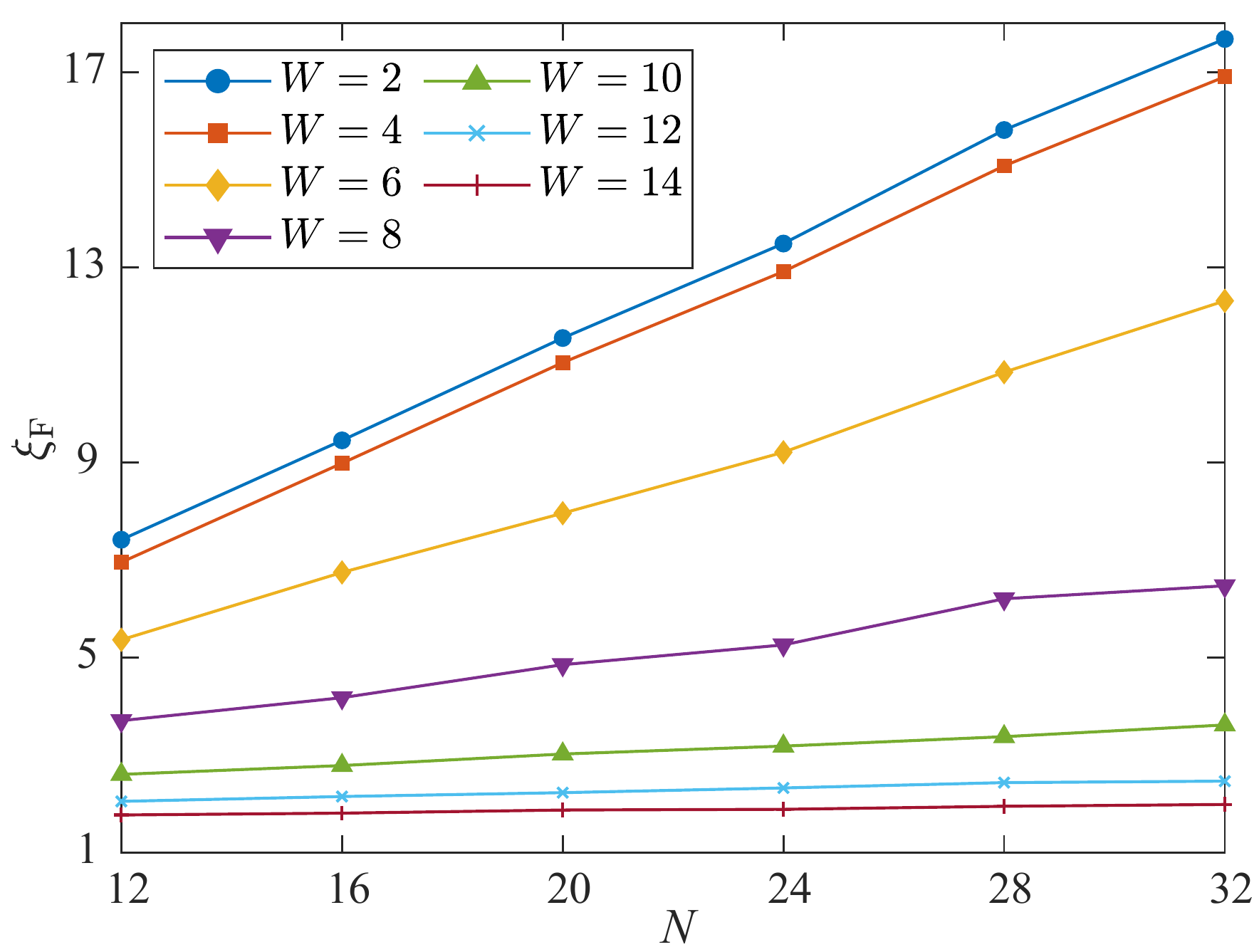}
	\caption{The characteristic localization length $\xi_{\textrm{F}}$ of the Floquet eigenstates versus the system size $N$ for different disorder strengths. The results are averaged over 1200 disorder realizations for $N\leq 16$, over 800 disorder realizations for $24\geq N>16$, and over 400 disorder realizations for $N>24$. The parameter used in the simulation is $\hbar \omega=0.05\delta_0$. }\label{Fig5}
\end{figure}

\begin{figure*}[!tb]
	\centering
	\includegraphics[width=1\linewidth]{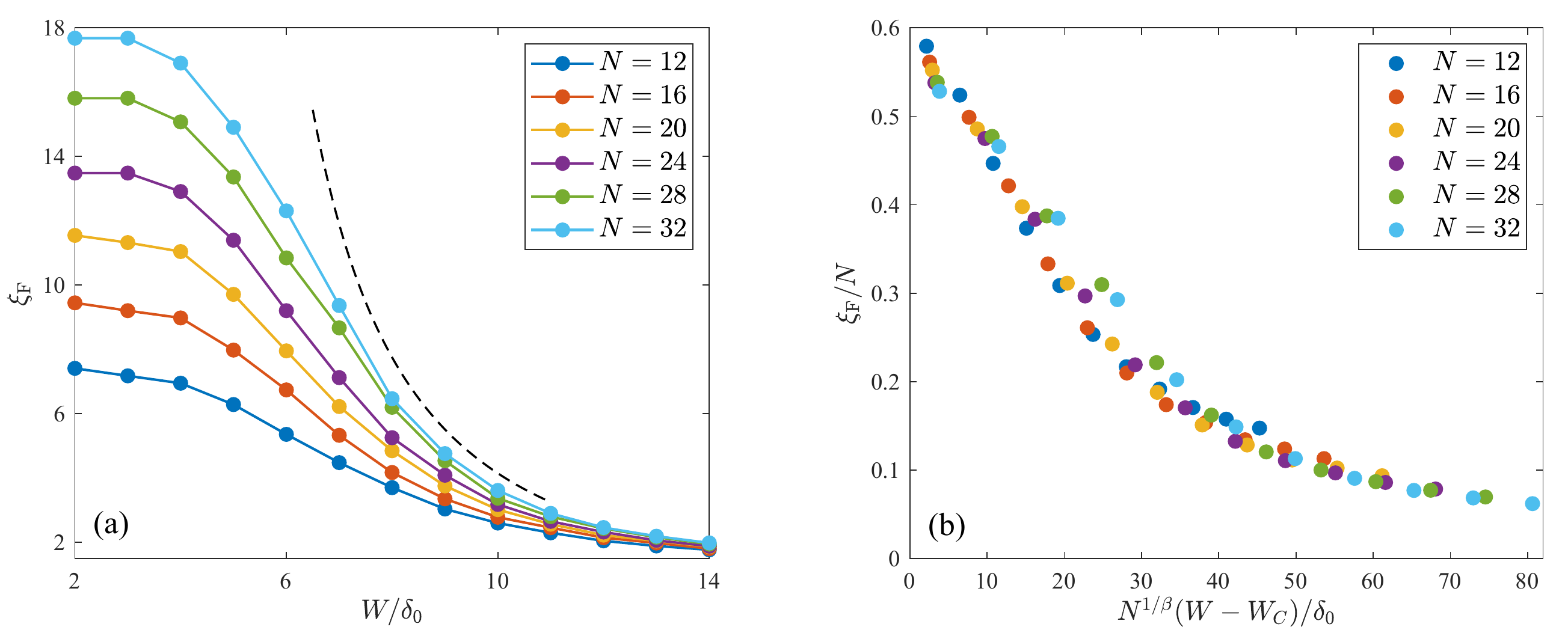}
	\caption{(a) Characteristic localization length $\xi_{\textrm{F}} $ of the Floquet eigenstates as a function of disorder strength W with different system sizes $N$. The curve of the dashed line is proportional to $(W-W_C)^{-\beta}$. (b) The collapse of $\xi_{\textrm{F}}/N$, plotted as a function of $N^{1/\beta}(W-W_C)/\delta_0$,  with $W_C\simeq3.1\delta_0$ and $\beta\simeq1.8$. The parameter used in the simulation is $\hbar \omega=0.05\delta_0$.}\label{Fig6}
\end{figure*}

\begin{figure*}[!tb]	
	\centering
	\includegraphics[width=1\linewidth]{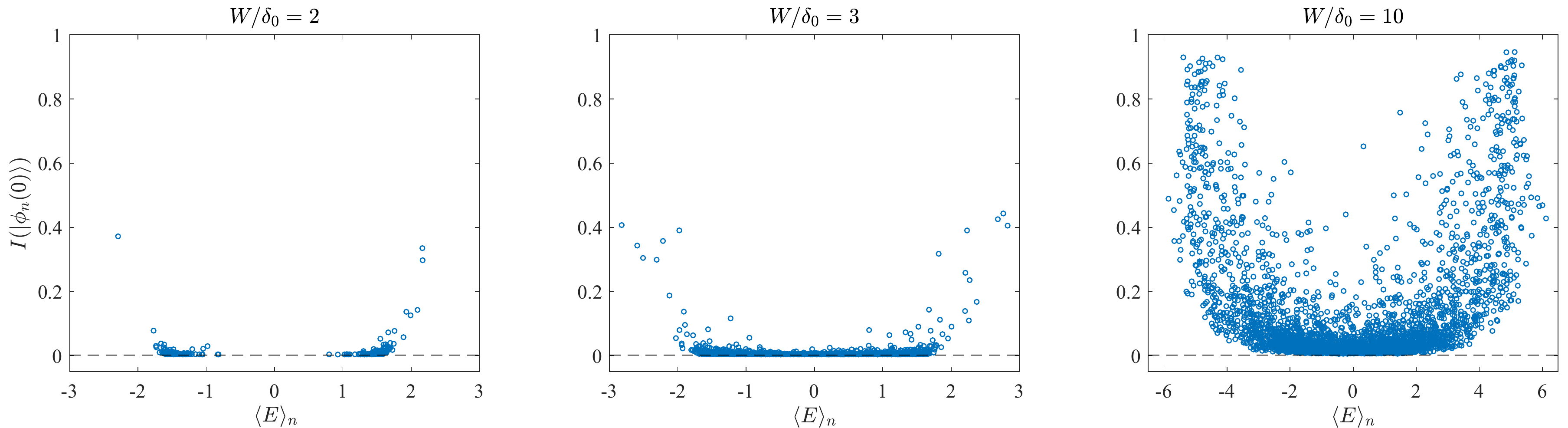}
	\caption{IPR $I(\ket{\phi_n(0)})$  versus the  time-averaged energy $\langle E \rangle_n$ of the corresponding Floquet eigenstate. The dashed lines indicate the value $1/N^2$ for totally delocalized states. The results are simulated by several disorder realizations with $N=24$ and $\hbar \omega=0.02\delta_0$.}\label{Fig7}
\end{figure*}

To explore why the quantized charge transport survives even though the instantaneous eigenstates  remain localized for the small disorder strength, we study the localization properties of the Floquet eigenstates. For the periodic driving, we write the  evolution operator over a pumping cycle in the Floquet  representation\cite{GRIFONI1998229} 
\begin{align}\label{Eq14}
\hat{U}(T, 0)=\sum_n e^{-i \mathcal{E}_n T / \hbar}\left|\phi_n(0)\right\rangle\left\langle\phi_n(0)\right|,
\end{align}
where $\ket{\phi_n(0)}$ is the Floquet eigenstate,   and $\varepsilon_n $ is the corresponding quasienergy with $-\pi / T \leq \varepsilon_n<\pi / T$. The characteristic localization length of the Floquet eigenstates is defined as 
\begin{align}\label{Eq15}
	\xi_{\textrm{F}}=\{\min _{n}\bar{I}(\ket{\phi_n(0)})\}^{-1/2},
\end{align}
where $I(\ket{\phi_n(0)}) = \sum_{j}|\bra{j} \ket{\phi_n(0)}  |^4.$  

Figure \ref{Fig5} shows the $\xi_{\textrm{F}}$ as a function of the lattice size $N$ for different disorder strengths $W$. For $W<4\delta_0$, $\xi_{\textrm{F}}\sim N$, which shows  the longer localization length than $\xi_{\textrm{I}}$ in Fig.~\ref{Fig4}. The Floquet eigenstates remain delocalized, where the periodic driving mixes the localized instantaneous eigenstates \cite{physRevLett.123.26601}, and  the quantized charge transport is thus survived for the small disorder strength despite the instantaneous eigenstates localized. However, for $W>10\delta_0$, the Floquet eigenstates are perfectly localized, where the charge transport is inhibited (see Fig.~\ref{Fig2}). This indicates that the disorder-induced transition from higher-order topological pumping to topologically trivial pumping is accompanied by the delocalization-localization transition of Floquent eigenstates.

To further determine the critical disorder strength $W_C$ of the disorder-induced phase transition for the driving system, we perform a finite-size scaling analysis  of $\xi_{\textrm{F}}$. As shown in Fig.~\ref{Fig6}(a), we infer that $\xi_{\textrm{F}}(W)\sim(W-W_C)^{-\beta}$ with $\beta\simeq1.8$ for  $W>W_C$ under the thermodynamic limit, where $W_C\simeq3.1\delta_0$ is the  critical disorder strength. By rescaling the data and plotting $\xi_{\textrm{F}}/N $ as a function of $N^{1/\beta}(W-W_C)/\delta_0$ in Fig.~\ref{Fig6}(b), we  observe the collapse of $\xi_{\textrm{F}}/N $. While, for $W<W_C$, $\xi_{\textrm{F}}$ scales as $N$ (see also Fig.~\ref{Fig5}). These indicate a  delocalization-localization phase transition. Moreover, the critical disorder strength $W_C$ extracted by this scaling analysis is compatible with the breaking of quantized charge transport in Fig.~\ref{Fig2}.

As shown in Fig.~\ref{Fig7}, we plot the energy-resolved IPR $I(\ket{\phi_n(0)})$ of Floquet eigenstates  over the time-averaged energy of Floquet eigenstates $\langle E \rangle_n=\frac{1}{T} \int_0^T d t\left\langle\phi_n(t)|\mathcal{H}_{\textrm{tot}}(t)| \phi_n(t)\right\rangle$ under CPBC. For $W=2\delta_0<W_C$, the time-averaged energy bands are separated by a gap, and nearly all the Floquet eigenstates are delocalized, leading to the quantized corner-to-corner charge transport at half filling. As the disorder rises, the energy gap decreases. Near the critical region with $W \simeq W_C$, the disorder makes two energy-band regions touch and merge, and breaks the quantized charge pumping, although most of Floquet eigenstates are still extended states. For the strong disorder with $W = 10 \delta_0$, most of the Floquet eigenstates are localized, inhibiting the charge transport.

\section{\label{sec:Conclusion}SUMMARY AND Conclusion}

We have investigated the higher-order Thouless pumping of the noninteracting fermionic BBH model in the presence of random onsite disorder.  We started with the corner-to-corner charge transport in the pumped clean system, where its charge flow is related to four higher-order Zak phases defined in corner periodic boundary conditions, and the quantized pump is characterized by the winding of higher-order Zak phase  (i.e., Chern number). When the random onsite disorder is introduced into the system, the instantaneous eigenstates remain localized. However, the quantized corner-to-corner charge transport is survived in the weak disorder, although it is broken as the disorder strength increases. The transition from higher-order topological pumping to trivial phase is characterized by the disorder-averaged Chern number (i.e., winding of higher-order Zak phase). Furthermore, we have shown that the quantized charge transport, in the weak disorder, is related to the extended Floquet states, where the periodic driving mixes localized instantaneous eigenstates. By analyzing the inverse participation ratio, we found that the phase transition from the quantized charge transport to topologically trivial pumping is accompanied by the disorder-induced delocalization-localization transition of Floquet states.

\begin{acknowledgments}
	T.L. acknowledges the support from National Natural Science Foundation of China (Grant No.~12274142), the Startup Grant of South China University of Technology (Grant No.~20210012) and Introduced Innovative Team Project of Guangdong Pearl River Talents Program (Grant No.~2021ZT09Z109). Q.A. thanks the support from Beijing Natural Science Foundation (Grant No.~1202017) and Beijing Normal University (Grant No.~2022129). M.Z. acknowledges the support from the National Natural Science Foundation of China (Grant No.~11475021) and the National Key Basic Research Program of China (Grant No.~2013CB922000). H.B.W. thanks the support from National Natural Science Foundation of China (Grant No.~61675028) and National Natural Science Foundation of China (Grant No.~12274037)
\end{acknowledgments}

\end{document}